\documentclass[12pt,showpacs,preprintnumbers,superscriptaddress,amsmath,amssymb,nofootinbib]{revtex4}
\usepackage{amsmath, graphicx}
\usepackage{dcolumn}
\usepackage{bm}
\usepackage{epstopdf}
\begin{document}
\newcommand{\hs}{\hspace*{0.5cm}}
\newcommand{\vs}{\vspace*{0.5cm}}
\newcommand{\be}{\begin{equation}}
\newcommand{\ee}{\end{equation}}
\newcommand{\bea}{\begin{eqnarray}}
\newcommand{\eea}{\end{eqnarray}}
\newcommand{\ben}{\begin{enumerate}}
\newcommand{\een}{\end{enumerate}}
\newcommand{\bde}{\begin{widetext}}
\newcommand{\ede}{\end{widetext}}
\newcommand{\nn}{\nonumber}
\newcommand{\crn}{\nonumber \\}
\newcommand{\Tr}{\mathrm{Tr}}
\newcommand{\non}{\nonumber}
\newcommand{\noi}{\noindent}
\newcommand{\al}{\alpha}
\newcommand{\la}{\lambda}
\newcommand{\bet}{\beta}
\newcommand{\ga}{\gamma}
\newcommand{\va}{\varphi}
\newcommand{\om}{\omega}
\newcommand{\pa}{\partial}
\newcommand{\+}{\dagger}
\newcommand{\fr}{\frac}
\newcommand{\sq}{\sqrt}
\newcommand{\bc}{\begin{center}}
\newcommand{\ec}{\end{center}}
\newcommand{\Ga}{\Gamma}
\newcommand{\de}{\delta}
\newcommand{\De}{\Delta}
\newcommand{\ep}{\epsilon}
\newcommand{\varep}{\varepsilon}
\newcommand{\ka}{\kappa}
\newcommand{\La}{\Lambda}
\newcommand{\si}{\sigma}
\newcommand{\Si}{\Sigma}
\newcommand{\ta}{\tau}
\newcommand{\up}{\upsilon}
\newcommand{\Up}{\Upsilon}
\newcommand{\ze}{\zeta}
\newcommand{\ps}{\psi}
\newcommand{\Ps}{\Psi}
\newcommand{\ph}{\phi}
\newcommand{\vph}{\varphi}
\newcommand{\Ph}{\Phi}
\newcommand{\Om}{\Omega}

\title{Lepton mixing and CP violation phase in the 3-3-1 model with neutral leptons based on $T_{13}$ flavor symmetry}
\author{Vo Van Vien}
\email{wvienk16@gmail.com} \affiliation{Department of Physics, Tay
Nguyen University, 567 Le Duan, Buon Ma Thuot, DakLak, Vietnam}


\begin{abstract}
We study a 3-3-1 model based on non-Abelian discrete symmetry group $T_{13}$ which accommodates lepton mixing with non-zero $\theta_{13}$ and CP violation phase. The neutrinos get
small masses and mixing with CP violation phase from $SU(3)_L$ antisextets which are all in triplets under $T_{13}$. If both breakings $T_{13}\rightarrow Z_3$ and $Z_{3}\rightarrow \{\mathrm{Identity}\}$ are taken palce in neutrino sector and $T_{13}$ is broken into $Z_3$ in lepton sector, the realistic neutrino mixing form is
obtained as a natural consequence of $P_l$ and $T_{13}$ symmetries. The model predicts the lepton mixing with non-zero $\theta_{13}$, and also gives a remarkable prediction of Dirac CP violation $\delta_{CP}=292.5^\circ$ in the normal spectrum, and $\delta_{CP}=303.161^\circ$ in the inverted spectrum which is still missing in the neutrino mixing matrix. There exist some regions of model parameters that can fit the experimental
data in 2014 on neutrino masses and mixing without perturbation. 

\end{abstract}

\maketitle

\section{\label{intro}Introduction}

\hs The Standard Model (SM) is one of the most successful and thoroughly tested theories in the elementary particle physics field, however, the origin of flavor structure, masses and mixings between generations of matter particles are unknown yet. Many experiments show that neutrinos have tiny masses and their mixing is sill mysterious \cite{altar1, altar2}. The neutrino mass and mixing is one of the most important evidence of beyond Standard Model physics. Among the possible extensions of SM, the 3-3-1 models, which encompass a class of models based on the gauge group  $\mbox{SU}(3)_C\otimes \mbox{SU}(3)_L \otimes
\mbox{U}(1)_X$ \cite{331m1,331m2,331m3,331m4,331m5, 331r1, 331r2,
331r3, 331r4, 331r5, 331r6, e3311, e3312,e331v1, e331v2, e3313, e3314}, have revealed interesting features.
 The first one is that the requirement of
anomaly cancelation together with that of asymptotic freedom of
QCD implies that the number of generations must necessarily be
equal to the number of colors, hence giving an explanation for the
existence of three generations. Furthermore, quark generations
should transform differently under the action of $SU(3)_L$. In
particular, two quark generations  should transform as triplets,
one as an antitriplet.

A fundamental relation holds among some of the generators of the
subgroups in the models \cite{T76, T77}: \be
Q=T_3+\beta T_8+X \label{Qoper}\ee
where $Q$ indicates the electric
charge, $T_3$ and $T_8$ are two of the $SU(3)$ generators and $X$
is the generator of $U(1)_X$. $\beta$ is a key parameter that
defines a specific variant of the model.

There are two typical variants of the 3-3-1 models as far as
lepton sectors are concerned. In the minimal version, three
$\mathrm{SU}(3)_L$ lepton triplets are $(\nu_L,l_L,l^c_R)$, where
$l_{R}$ are ordinary right-handed charged-leptons \cite{331m1,
331m2, 331m3, 331m4, 331m5}   . In the second version, the third
components of lepton triplets are the  right-handed neutrinos,
$(\nu_L,l_L,\nu^c_R)$ \cite{331r1,331r2,331r3,331r4,331r5,331r6}.
To have a model with the realistic neutrino mixing matrix, we
should consider another variant of the form $(\nu_L,l_L,N^c_R)$
where $N_R$ are three new fermion singlets under SM
symmetry with vanishing lepton-numbers \cite{dlshA4, dlsvS4,
dlnvS3, vlD4}   .

The recent data have provided the evidence for a non-vanishing value of the smallest mixing angle $\theta_{13}$ \cite{smirnov}
but the tri-bimaximal form for explaining the lepton mixing scheme was first proposed by Harrison-Perkins-Scott (HPS), which
apart from the phase redefinitions, is given by \cite{hps1,hps2,hps3,hps4}
\begin{eqnarray}
U_{\mathrm{HPS}}=\left(
\begin{array}{ccc}
\frac{2}{\sqrt{6}}       &\frac{1}{\sqrt{3}}  &0\\
-\frac{1}{\sqrt{6}}      &\frac{1}{\sqrt{3}}  &\frac{1}{\sqrt{2}}\\
-\frac{1}{\sqrt{6}}      &\frac{1}{\sqrt{3}}  &-\frac{1}{\sqrt{2}}
\end{array}\right),\label{Uhps}
\end{eqnarray}
can be considered  as a good approximation for the recent
neutrino experimental data.
In fact, the absolute
values of the entries of the lepton mixing matrix $U_{PMNS}$ approximately are given by \cite{Uij}
\be
\left|U_{\mathrm{PMNS}}\right|=\left(
\begin{array}{ccc}
0.795-0.846     &0.513-0.585  &0.126-0.178\\
0.205-0.543     &0.416-0.730  &0.579-0.808\\
0.215-0.548     &0.409-0.725  &0.567-0.800
\end{array}\right),\label{Uij}
\ee
while the most recent data in PDG2014 \cite{PDG2014} imply\footnote{In this paper, NH and IH stand for the normal
and inverted mass hierarchies, respectively.}:
\bea
&&\sin^2(2\theta_{12})=0.846\pm 0.021 , \,\, \Delta m^2_{21}=(7.53\pm 0.18)\times 10^{-5} \mathrm{eV}^2,\crn
&& \sin^2(2\theta_{13})=(9.3\pm0.8)\times 10^{-2},\crn
&& \sin^2(2\theta_{23})=0.999^{+0.001}_{-0.018}, \,\,\,  \Delta m^2_{32}=(2.44\pm0.06)\times
10^{-3}\mathrm{eV}^2 ,\, (\mathrm{NH}),\label{PDG2014}\\
&&\sin^2(2\theta_{23})=1.000^{+0.000}_{-0.017}, \,\,\,  \Delta m^2_{32}=(2.52\pm 0.07)\times
10^{-3}\mathrm{eV}^2 ,\,\, (\mathrm{IH}),\nn\eea
with a slight deviation from Tri-bimaximal mixing form given in (\ref{Uhps}). 
These large neutrino mixing angles are completely different from the quark
 mixing ones defined by the Cabibbo- Kobayashi-Maskawa (CKM) matrix \cite{CKM, CKM1}   .
 This has stimulated work on flavor symmetries and non-Abelian discrete
 symmetries are considered to be the most attractive candidate to
formulate dynamical principles that can lead to the flavor
mixing patterns for quarks, lepton and some other issues related to the flavor physics. In order to overcome these problems, plenty of models based on the principle of symmetry, flavor symmetry, have been discussed. Among them, non-Abelian discrete symmetries are well discussed as plausible possibilities. In particular, the fact that the lepton mixing matrix ($U_{PMNS}$) shows very good agreement with the tri-bimaximal form \cite{hps1,hps2,hps3,hps4} implies that flavor structure
is originated from a symmetry.

There are many recent models based on
the non-Abelian discrete symmetries, such as $A_4$ ~\cite{A41, A42, A43, A44, A45,A46, A47, A48, A49, A410, A411, A412, A413, A414, A415, A416, A417, A418}  , $A_5$ \cite{A51, A52, A53, A54, A55, A56, A57, A58, A59, A510, A511, A512, A513} ,
$S_3$\cite{S31,S32,S33,S34,S35,S36,S37,S38,S39,S310,S311,S312,S313,S314,S315,S316,S317,S318,S319,S320,S321,S322,S323,S324,S325,S326,S327,S328,S329,S330,S331,S332,S333,S334,S335,S336,S337,S338,S339,S340,S341,S342} , $S_4$ \cite{S41,S42,S43,S44,S45,S46,S47,S48,S49,S410,S411,S412,S413,S414,S415,S416,S417,S418,S419,S420,S421,S422,S423,S424,S425,S426,S427,S428,S429} , $D_4$ \cite{D41,D42,D43,D44,D45,D46,D47,D48,D49,D410,D411,D412}   , $D_5$ \cite{D51, D52}   ,
$T'$ \cite{Tp1,Tp2,Tp3,Tp4,Tp5} and so forth. In the context of the 3-3-1 model, the non-abelian discrete symmetries $A_4, S_3, S_4, D_4, T_7$
have been explored \cite{dlshA4, dlsvS4, dlnvS3, vlD4,vlS4,vlS3,vlT7}, however, $T_{13}$ symmetry has not been considered before in this kind of the model. For the similar works on $T_{13}$, let us call the reader's attention to Refs. \cite{T131, T132, T133}. 
 In Ref. \cite{dlsvS4} we have studied the 3-3-1 model with neutral fermions based
    on $S_4$ group, in which most of the Higgs multiplets are in triplets under $S_4$
    except $\chi$ lying  in a singlet, and the exact tribimaximal form \cite{hps1,hps2,hps3,hps4} is obtained.
As we know, the recent considerations have implied $\theta_{13}\neq 0$ \cite{A41, A42, A43, A44, A45,A46, A47, A48, A49, A410, A411, A412, A413, A414, A415, A416, A417, A418, S31,S32,S33,S34,S35,S36,S37,S38,S39,S310,S311,S312,S313,S314,S315,S316,S317,S318,S319,S320,S321,S322,S323,S324,S325,S326,S327,S328,S329,S330,S331,S332,S333,S334,S335,S336,S337,S338,S339,S340,S341,S342, S41,S42,S43,S44,S45,S46,S47,S48,S49,S410,S411,S412,S413,S414,S415,S416,S417,S418,S419,S420,S421,S422,S423,S424,S425,S426,S427,S428,S429}, but small as given in (\ref{PDG2014}).
     This problem has been improved in Refs. \cite{dlnvS3, vlD4} by adding new Higgs multiplets and using the perturbation theory up to the first order.

CP violation plays a crucial role in our understanding of the
observed baryon asymmetry of the Universe (BAU) \cite{CPvio}. In the SM, CP symmetry is violated due
to a complex phase in the CKM matrix \cite{CKM, CKM1}. However,
since the extent
 of CP violation in the SM is not enough for achieving the observed BAU,  we need new source of CP violation for successful BAU.
On the other hand, CP violations in the lepton sector are
imperative if the BAU could be realized through leptogenesis. So,
any hint or observation of the leptonic CP violation  can
strengthen our belief in leptogenesis \cite{CPvio}. The violation of the CP symmetry is a very important ingredient of any
dynamical mechanism which intends to explain both low energy CP
violation and the baryon asymmetry. Renormalizable gauge theories
are based
 on the spontaneous symmetry breaking mechanism, and it is natural to have the spontaneous CP violation
  as an integral part of that mechanism. Determining all
possible sources of CP violation is a fundamental challenge for high energy physics. In theoretical
 and economical viewpoints, the spontaneous CP breaking necessary to generate the baryon asymmetry
  and leptonic CP violation at low energies brings us to a common source which comes from the
  phase of the scalar field responsible for the spontaneous CP breaking at a high energy scale \cite{CPvio}.

In this work, we investigate
another choice with the Frobenius group $T_{13}$, which is isomorphic to $Z_{13}\rtimes Z_3$. The  $T_{13}$ group is a subgroup of $SU(3)_L$, and known as the minimal non-Abelian discrete group having two complex triplets as the irreducible representations.  $T_{13}$ contains two complex irreducible representations $\underline{3}_1, \underline{3}_2$ and three singlets $\underline{1}_0$, $\underline{1}_1$, $\underline{1}_2$ . This feature is useful to
separate the three families of fermions from the others as requirement of the $3-3-1$ models. Namely, in this work, three
inequivalent singlet representations $\underline{1}_0,
\underline{1}_1$ and $\underline{1}_2$  of $T_{13}$ play
a crucial role in consistently reproducing fermion masses and
mixings which allow to naturally accommodate the three right-handed charged-leptons, three left-handed components of the ordinary quarks and the right-handed components of the exotic quarks. On the other hand, three left- handed leptons and three right-handed components of the ordinary up-quarks and down-quarks are accommodated in two complex irreducible representations which can generate small masses for neutrinos since the good feature of tensor products $\underline{\bar{3}}_{1}\otimes \underline{3}_{1}$ under $T_{13}$: $\underline{\bar{3}}_{1}\otimes \underline{3}_{1}=1_0\oplus 1_1 \oplus 1_2\oplus \underline{3}_{2}\oplus \underline{\bar{3}}_{2}$.

A brief of the theory of $T_{13}$
group is given in Appendix \ref{apa}. Two important tensor products of $T_{13}$ used to construct the Yukawa interactions that responsible for fermion masses are
\bea
\underline{\bar{3}}_{1}\otimes \underline{3}_{1}&=&\underline{3}_{1}\otimes \underline{\bar{3}}_{1}=1_0(11+22+33)\oplus 1_1(11+\om22+\om^233)\crn
&\oplus &1_2(11+\om^222+\om 33)\oplus \underline{3}_{2}(21,32,13)\oplus \underline{\bar{3}}_{2}(12,23,31), \label{important}\\
\underline{\bar{3}}_{2}\otimes \underline{3}_{2}&=&\underline{3}_{2}\otimes \underline{\bar{3}}_{2}=1_0(11+22+33)\oplus 1_1(11+\om22+\om^233)\crn
&\oplus &1_2(11+\om^222+\om 33)\oplus \underline{3}_{1}(23,31,12)\oplus \underline{\bar{3}}_{1}(32,13,21).\nn
\eea

The rest of this work is organized
as follows. In Sec. \ref{fermion} and \ref{Chargedlep} we present
the necessary elements of the 3-3-1 model with $T_{13}$ symmetry
as well as introducing necessary Higgs fields responsible for the
charged lepton masses. In Sec. \ref{quark}, we  discuss on quark
sector. Sec. \ref{neutrino} is devoted for the neutrino mass and
mixing. We summarize our results and make conclusions in the
section \ref{conclus}. Appendix \ref{apa} presents a brief of the
$T_{13}$ theory. 

\section{Fermion content\label{fermion}}

In the model under consideration, the gauge symmetry is given by $\mathrm{SU}(3)_C\otimes
\mathrm{SU}(3)_L \otimes \mathrm{U}(1)_X$, where the electroweak
factor $\mathrm{SU}(3)_L \otimes \mathrm{U}(1)_X$ is extended from
those of the SM while the strong interaction sector is retained.
Each lepton family includes a new lepton singlet carrying
no lepton-number $(N_R)$ and is arranged under the
$\mathrm{SU}(3)_L$ symmetry as a triplet $(\nu_L, l_L, N^c_R)$ and
a singlet $l_R$. We consider the version with $\beta=-\fr{1}{\sqrt{3}}$ in Eq.(\ref{Qoper}), hence, there is no exotic electric charges in the fundamental fermion, scalar and adjoint
gauge boson representations. Since the particles in the lepton triplet have different lepton
numbers (0 and 1), so the lepton number in the model  does not
commute with the gauge symmetry unlike the SM. Therefore, it is
better to work with a new conserved charge $\mathcal{L}$ commuting
with the gauge symmetry and related to the ordinary lepton number ($L$)
by diagonal matrices \cite{dlshA4,dlsvS4, dlnvS3, vlD4, clong}
 \bea L=\fr{2}{\sqrt{3}}T_8+\mathcal{L}.\label{LLcong}\eea
The lepton charge arranged in this way [i.e. $L(N_R)=0$ as
assumed] is in order to prevent unwanted interactions due to
$\mathrm{U}(1)_\mathcal{L}$ symmetry and breaking to obtain the consistent lepton and
quark spectra. By this embedding, exotic quarks $U, D$ as well as
new non-Hermitian gauge bosons $X^0$, $Y^\pm$ possess lepton
charges as of the ordinary leptons, $L(D)=-L(U)=L(X^0)=L(Y^{-})=1$.  

In the model under consideration, 
we put three left- handed leptons, three right-handed components of the ordinary up-quarks in $\underline{3}_1$ and three right-handed components of the ordinary down-quarks in $\underline{\bar{3}}_1$, while three right-handed charged-leptons, three left-handed components of the ordinary quarks and three right-handed components of the exotic quarks are in the singlets. 
Under the $[\mathrm{SU}(3)_L, \mathrm{U}(1)_X,
\mathrm{U}(1)_\mathcal{L},\underline{T}_{13}]$ symmetries as
proposed, the fermions of the model, respectively, transform as follows \bea
\psi_{L} &\equiv&\psi_{1,2,3L}= \left( \nu_{1,2,3L} \,\,\,\,\,
    l_{1,2,3L} \,\,\,\,\,
    N^c_{1,2,3R}\right)^T\sim [3,-1/3,2/3,\underline{3}_1],\crn
l_{1R}&\sim&[1,-1,1,\underline{1}_0],\,\,\,
l_{2 R}\sim[1,-1,1,\underline{1}_1],\,\,\, l_{3 R}\sim[1,-1,1,\underline{1}_2] , \crn
 Q_{1 L}&\equiv &
 \left( d_{1 L} \hs
  -u_{1 L}  \hs
    D_{1 L}\right)^T\sim[3^*,0,1/3,\underline{1}_1], \crn
    Q_{2 L}&\equiv &
 \left( d_{2 L} \hs
  -u_{2 L}  \hs
    D_{2 L}\right)^T\sim[3^*,0,1/3,\underline{1}_2], \label{Fermion}\\
    Q_{3L}&=& \left(u_{3L} \hs
    d_{3L} \hs
    U_{L} \right)^T\sim[3,1/3,-1/3,\underline{1}_0],\crn
u_{i R}&\sim& [1,2/3,0,\underline{3}_1],\hs d_{i R}\sim [1,-1/3,0,\underline{\bar{3}}_1], \crn
D_{1 R}&\sim& [1,-1/3,1,\underline{1}_2],\,\, D_{2 R}
\sim[1,-1/3,1,\underline{1}_1], \,\, U_R \sim[1,2/3,-1,\underline{1}_0].\nn\eea where the
subscript numbers on fields indicate to respective families which
also  in order define components of their $T_{13}$ multiplets. In the following, we consider
possibilities of generating the masses for the fermions. The
scalar multiplets needed for the purpose are also introduced.

\section{Charged-lepton masses \label{Chargedlep}}
The charged-lepton masses arise from the couplings of $\bar{\psi}_{L} l_{iR}\,\,(i=1,2,3)$ to scalars,
 where $\bar{\psi}^c_{L} \psi_{L}$ transforms as $3^*\oplus 6$ under
$\mathrm{SU}(3)_L$ and $\underline{3}_1$  under $T_{13}$. To generate masses for the charged leptons, we need one $SU(3)_L$ Higgs triplets put in $\underline{3}_1$ under $T_{13}$, 
\bea \phi = \left( \phi^+_1 \hs  \phi^0_2 \hs  \phi^+_3 \right)^T\sim [3,2/3,-1/3, \underline{3}_1].\eea
The Yukawa interactions are
 \bea -\mathcal{L}_{l}&=&h_1 (\bar{\psi}_{L}
\phi)_{\underline{1}_0} l_{1R}+h_2 (\bar{\psi}_{L}\phi)_{\underline{1}_2}l_{2 R}
+h_3 (\bar{\psi}_{L}\phi)_{\underline{1}_1} l_{3 R}+H.c\crn
&=&h_1 (\bar{\psi}_{1L}
\phi_1+\bar{\psi}_{2L}
\phi_2+\bar{\psi}_{3L}
\phi_3) l_{1R}\crn
&+&h_2(\bar{\psi}_{1L}
\phi_1+\om^2 \bar{\psi}_{2L}
\phi_2+\om\bar{\psi}_{3L}
\phi_3) l_{2 R}\crn
 &+&h_3(\bar{\psi}_{1L}
\phi_1+\om \bar{\psi}_{2L}
\phi_2+\om^2\bar{\psi}_{3L}
\phi_3) l_{3 R}+H.c. \eea
Following the potential minimization
condition, we have the followings alignments:
\begin{itemize}
\item[(1)] The first alignment: $\langle \phi_1\rangle\neq\langle \phi_2\rangle\neq\langle \phi_3\rangle$ then $T_{13}$ is completely broken.
\item[(2)] The second alignment: $0\neq\langle \phi_1\rangle\neq\langle \phi_2\rangle=\langle\phi_3\rangle\neq0$ or $0\neq\langle \phi_1\rangle=\langle \phi_3\rangle\neq\langle \phi_2\rangle\neq0$ or $0\neq\langle \phi_1\rangle=\langle \phi_2\rangle\neq\langle \phi_3\rangle\neq0$ then $T_{13}$ is completely broken.
\item[(3)] The third alignment: $\langle \phi_1\rangle=\langle
\phi_2\rangle =\langle \phi_3\rangle \neq 0$ then $T_{13}$ is  broken into $Z_3$ that consists of the elements  \{$e, b, b^2$\}.
\item[(4)] The fourth alignment: $0=\langle \phi_1\rangle\neq\langle \phi_2\rangle\neq\langle \phi_3\rangle \neq 0$ or $0=\langle \phi_2\rangle\neq\langle \phi_1\rangle\neq\langle \phi_3\rangle \neq 0$ or $0\neq \langle \phi_1\rangle\neq\langle \phi_2\rangle\neq\langle \phi_3\rangle=0$ then $T_{13}$ is completely broken.
\item[(5)] The fifth alignment: $0=\langle \phi_1\rangle\neq\langle \phi_2\rangle=\langle \phi_3\rangle \neq 0$ or $0=\langle \phi_2\rangle\neq\langle \phi_1\rangle=\langle \phi_3\rangle \neq 0$ or $0\neq \langle \phi_1\rangle=\langle\phi_2\rangle\neq\langle \phi_3\rangle=0$ then $T_{13}$ is completely broken.
\item[(6)] The sixth alignment: $0=\langle \phi_1\rangle=\langle \phi_2\rangle\neq\langle \phi_3\rangle$ or $0=\langle \phi_1\rangle=\langle \phi_3\rangle\neq\langle \phi_2\rangle$ or $0\neq \langle \phi_1\rangle\neq \langle\phi_2\rangle=\langle \phi_3\rangle=0$ then $T_{13}$ is completely broken.
\end{itemize}
Theoretically, a possibility that tribimaximal mixing matrix $U_{HPS}$ in (\ref{Uhps}) can be decomposed into only two independent rotations
may provide a hint for some underlying structure in the lepton sector, such as
\begin{eqnarray}
U_{\mathrm{HPS}}&=&\left(
\begin{array}{ccc}
\frac{2}{\sqrt{6}}       &\frac{1}{\sqrt{3}}  &0\\
-\frac{1}{\sqrt{6}}      &\frac{1}{\sqrt{3}}  &\frac{1}{\sqrt{2}}\\
-\frac{1}{\sqrt{6}}      &\frac{1}{\sqrt{3}}  &-\frac{1}{\sqrt{2}}
\end{array}\right)
=\frac{1}{\sqrt{3}}\left(
\begin{array}{ccc}
1       &1  &1\\
1     &\om  &\om^2\\
1    &\om^2  &\om \end{array}\right)\left(
\begin{array}{ccc}
0      &1 &0\\
\frac{1}{\sqrt{2}}      &0 &\frac{i}{\sqrt{2}}\\
\frac{1}{\sqrt{2}}      &0 &-\frac{i}{\sqrt{2}}
\end{array}\right)\cong U^+_L U_{\nu},\label{Ulepdecomp}
\end{eqnarray}
where $\omega = \exp(2 \pi i/3) = -1/2 + i\sqrt{3}/2$.

To obtain charged - lepton mixing satisfying (\ref{Ulepdecomp}), in this work we
impose only the breaking $T_{13}\rightarrow Z_3$ in charged lepton sector, and this happens with the third alignment as above, i.e, $\langle \phi\rangle=(\langle \phi_1\rangle,\langle \phi_1\rangle,\langle \phi_1\rangle )$ under $T_{13}$, where
\bea
\langle \phi_1\rangle=(0\hs v \hs 0)^T.\label{vevphi}
\eea
The mass Lagrangian of the
charged leptons takes the form:
 \bea -\mathcal{L}^{mass}_{l}&=&h_1v \bar{l}_{1L}l_{1R}+h_2v\bar{l}_{1L} l_{2 R}+h_3v\bar{l}_{1L}l_{3 R}\crn
 &+&h_1v\bar{l}_{2L}l_{1R}+h_2\om^2v \bar{l}_{2L} l_{2 R}+h_3\om v\bar{l}_{2L}l_{3 R}\crn
 &+&h_1v\bar{l}_{3L}l_{1R}+h_2\om v\bar{l}_{3L} l_{2 R}+h_3\om^2 v\bar{l}_{3L}l_{3 R}
 +H.c. \label{Llep}\eea
 We can rewite the Lagrangian (\ref{Llep}) in the matrix form as follows
\bea
-\mathcal{L}^{\mathrm{mass}}_l=(\bar{l}_{1L},\bar{l}_{2L},\bar{l}_{3L})
M_l (l_{1R},l_{2R},l_{3R})^T+H.c,\eea 
where \bea M_l=
\left(%
\begin{array}{ccc}
  h_1v & h_2v& h_3 v \\
   h_1v &\,\,\,\,\, h_2v\om^2 &\,\, h_3 v\om \\
  h_1v & \,\,\, h_2v\om &\,\,\,\, h_3 v\om^2 \\
\end{array}%
\right).\label{Mlep0}\eea
The mass matrix $M_l$ is then diagonalized, \bea U^\dagger_L M_lU_R=\left(%
\begin{array}{ccc}
  \sqrt{3}h_1 v & 0 & 0 \\
  0 & \sqrt{3}h_2 v& 0 \\
  0 & 0 & \sqrt{3}h_3 v\\
\end{array}%
\right)\equiv\left(%
\begin{array}{ccc}
  m_e & 0 & 0 \\
  0 & m_\mu & 0 \\
  0 & 0 & m_\tau \\
\end{array}%
\right),\eea where \bea U_L=\fr{1}{\sqrt{3}}\left(%
\begin{array}{ccc}
  1 & 1 & 1 \\
  1 & \om^2 & \om \\
  1 & \om & \om^2 \\
\end{array}%
\right),\hs U_R=1.\label{lep}\eea
The charged
lepton Yukawa couplings $h_{1, 2, 3}$ are defined as follows:
\bea h_1&=& \frac{m_e}{\sqrt{3}v},\hs  h_2= \frac{m_\mu}{\sqrt{3}v},\hs h_3= \frac{m_\tau}{\sqrt{3}v}.\label{h1h2h3}
 \eea
The experimental  values for masses of  the charged leptons are given in \cite{PDG2014}:
\bea m_e\simeq0.51099\, \textrm{MeV},\hs \ m_{\mu}=105.65837 \ \textrm{MeV},\hs m_{\tau}=1776.82\,
\textrm{MeV} \label{Lepmas}\eea
It follows that $h_1\ll h_2\ll h_3$. On the other hand, if we choose the VEV $v \sim 100 GeV$ then
\bea
h_1\sim 10^{-6},\,\,\, h_2\sim 10^{-4},\,\,\, h_3\sim 10^{-2},\label{hi}\eea
i.e, in
the model under consideration, the hierarchy between the masses for charged-leptons can be achieved if there exists a hierarchy between Yukawa couplings $h_i \, (i=1,2,3)$ in charged-lepton sector as given in (\ref{hi}).

\section{\label{quark}Quark masses}
To generate masses for quarks with a minimal Higgs content, we
additionally introduce the following Higgs triplets
\bea \chi&=&
\left(  \chi^0_1 \hs
  \chi^-_2 \hs
  \chi^0_3 \right)^T\sim[3,-1/3,2/3,\underline{1}_0],\crn
\eta&=&
\left( \eta^0_1 \hs
  \eta^-_2 \hs
  \eta^0_3\right)^T\sim[3,-1/3,-1/3,\underline{\bar{3}}_1].
  \label{chieta}\eea
The Yukawa interactions are
 \bea -\mathcal{L}_q &=&
f_3(\bar{Q}_{3L}\chi)_{\underline{1}_0}U_R + f_1 (\bar{Q}_{1L}\chi^*)_{\underline{1}_1}D_{1 R}+ f_2 (\bar{Q}_{2L}\chi^*)_{\underline{1}_2}D_{2 R}\crn
&+&h^d_1\bar{Q}_{1 L}(\eta^*d_{i R})_{\underline{1}_2}+h^d_2\bar{Q}_{2 L}(\eta^*d_{i R})_{\underline{1}_1}+h^d_{3}\bar{Q}_{3L}(\phi d_{iR})_{\underline{1}_0}\crn
 &+& h^u_1\bar{Q}_{1L}(\phi^* u_{i R})_{\underline{1}_2}+ h^u_2\bar{Q}_{2L}(\phi^* u_{i R})_{\underline{1}_1}+h^u_{3}
\bar{Q}_{3L}(\eta u_{iR})_{\underline{1}_0} +H.c.\crn
&=&f_3(\bar{Q}_{3L}\chi)_{\underline{1}_0}U_R + f_1 (\bar{Q}_{1L}\chi^*)_{\underline{1}_1}D_{1 R}+ f_2 (\bar{Q}_{2L}\chi^*)_{\underline{1}_2}D_{2 R}\crn
&+&h^d_1\bar{Q}_{1 L}(\eta^*_1d_{1 R}+\om^2\eta^*_2d_{2 R}+\om\eta^*_3d_{3 R})\crn
&+&h^d_2\bar{Q}_{2 L}(\eta^*_1d_{1 R}+\om\eta^*_2d_{2 R}+\om^2\eta^*_3d_{3 R})\crn
&+&h^d_{3}\bar{Q}_{3L}(\phi_1d_{1 R}+\phi_2d_{2 R}+\phi_3d_{3 R})\crn
 &+& h^u_1\bar{Q}_{1L}(\phi^*_1 u_{1 R}+\om^2\phi^*_2 u_{2 R}+\om\phi^*_3 u_{3 R})\crn
&+& h^u_2\bar{Q}_{2L}(\phi^*_1 u_{1 R}+\om\phi^*_2 u_{2 R}+\om^2\phi^*_3 u_{3 R})\crn
&+&h^u_{3}
\bar{Q}_{3L}(\eta_1 u_{1 R}+\eta_2 u_{2 R}+\eta_3 u_{3 R}) +H.c,\label{Lquarks}\eea
where a residual symmetry of lepton number $P_l\equiv (-1)^L$,
called ''lepton parity'' \cite{dlnvS3, dlshA4} has been introduced in order to
suppress the mixing between ordinary quarks and exotic quarks. In this
framework we assume that the lepton parity is an exact symmetry,
not spontaneously broken. This means that  due to the lepton
parity conservation, the fields carrying  lepton number ($L=\pm
1$) $\eta_3$ and $\chi_1$ cannot develop VEV.
Suppose that, under $T_{13}$, the VEVs of $\chi$ and $\eta$ are $\langle\chi\rangle$ and $\langle\eta\rangle=(\langle\eta_1\rangle,\langle\eta_1\rangle,\langle\eta_1\rangle)$, respectively, where
 \bea \langle\chi\rangle&=&
  (0 \hs
  0 \hs
  v_\chi)^T,\,\hs
\langle\eta_1\rangle=
( u \hs
  0 \hs
  0 )^T,\label{vevchieta}\eea
The quark mass Lagrangian are
\bea -\mathcal{L}^{mass}_q
&=&f_3 v_\chi \bar{U}_{L}U_R + f_1 v_\chi \bar{D}_{1L}D_{1 R}+f_2 v_\chi \bar{D}_{2L}D_{2 R}\crn
&+&h^d_1u\bar{d}_{1 L}d_{1 R}+\om^2h^d_1u\bar{d}_{1 L}d_{2 R}+\om h^d_1u\bar{d}_{1 L}d_{3 R}\crn
&+&h^d_2u \bar{d}_{2 L}d_{1 R}+\om h^d_2u \bar{d}_{2 L}d_{2 R}+\om^2 h^d_2u \bar{d}_{2 L}d_{3 R}\crn
&+&h^d_{3}v \bar{d}_{3L}d_{1 R}+h^d_{3}v \bar{d}_{3L} d_{2 R}+h^d_{3}v \bar{d}_{3L}d_{3 R}\crn
 &-& h^u_1v \bar{u}_{1L}u_{1 R}-\om^2h^u_1v \bar{u}_{1L}u_{2 R}-\om h^u_1v \bar{u}_{1L} u_{3 R}\crn
&-& h^u_2 v\bar{u}_{2L}u_{1 R}-\om h^u_2 v\bar{u}_{2L} u_{2 R}-\om^2\ h^u_2 v\bar{u}_{2L}u_{3 R}\crn
&+&h^u_{3}u \bar{u}_{3L} u_{1 R}+h^u_{3}u\bar{u}_{3L} u_{2 R}+h^u_{3}u\bar{u}_{3L}u_{3 R}+H.c.\label{Lquarks}\eea
The exotic quarks  get
masses \bea m_U=f_3 v_\chi,\hs m_{D_{1,2}}=f_{1,2} v_\chi, \eea and the mass
Lagrangian of the ordinary quarks reads: \bea
-\mathcal{L}^{mass}_q &=&h^d_1u\bar{d}_{1 L}d_{1 R}+\om^2h^d_1u\bar{d}_{1 L}d_{2 R}+\om h^d_1u\bar{d}_{1 L}d_{3 R}\crn
&+&h^d_2u \bar{d}_{2 L}d_{1 R}+\om h^d_2u \bar{d}_{2 L}d_{2 R}+\om^2 h^d_2u \bar{d}_{2 L}d_{3 R}\crn
&+&h^d_{3}v \bar{d}_{3L}d_{1 R}+h^d_{3}v \bar{d}_{3L} d_{2 R}+h^d_{3}v \bar{d}_{3L}d_{3 R}\crn
 &-& h^u_1v \bar{u}_{1L}u_{1 R}-\om^2h^u_1v \bar{u}_{1L}u_{2 R}-\om h^u_1v \bar{u}_{1L} u_{3 R}\crn
&-& h^u_2 v\bar{u}_{2L}u_{1 R}-\om h^u_2 v\bar{u}_{2L} u_{2 R}-\om^2\ h^u_2 v\bar{u}_{2L}u_{3 R}\crn
&+&h^u_{3}u \bar{u}_{3L} u_{1 R}+h^u_{3}u\bar{u}_{3L} u_{2 R}+h^u_{3}u\bar{u}_{3L}u_{3 R}+H.c.\crn
&=&(\bar{u}_{1L}\hs \bar{u}_{2L}\hs \bar{u}_{3L}) M_u (u_{1R}\hs  u_{2R}\hs u_{3R})^T\crn
&+&(\bar{d}_{1L}\hs  \bar{d}_{2L}\hs \bar{d}_{3L}) M_d (d_{1R}\hs d_{2R}\hs  d_{3R})^T+H.c. \label{LquarkD4exp}\eea
 From
(\ref{LquarkD4exp}), the mass matrices for the ordinary up-quarks
and down-quarks are, respectively, obtained as follows:
\bea M_u =
\left(%
\begin{array}{ccc}
  -h^u_1 v & -h^u_1 \om^2 v  & -h^u_1 \om v  \\
   -h^u_2 v & -h^u_2 \om v  & -h^u_2 \om^2 v   \\
  h^u_3 u & h^u_3 u & h^u_3 u \\
\end{array}%
\right),\hs M_d=
\left(%
\begin{array}{ccc}
  h^d_1 u & h^d_1 \om^2 u & h^d_1 \om u  \\
   h^d_2 u & h^d_2 \om u  & h^d_2 \om^2 u   \\
  h^d_3 v & h^d_3 v & h^d_3 v \\
\end{array}%
\right).\label{MuMd} \eea
The matrices $M_u, M_d$ in (\ref{MuMd}) are, respectively, diagonalized as
\bea V^{u+}_L M_u V^u_R&=&
\left(%
\begin{array}{ccc}
  -\sqrt{3}h^u_1 v & 0 & 0 \\
  0 & -\sqrt{3}h^u_2 v & 0 \\
  0 & 0 & \sqrt{3}h^u_3 u \\
\end{array}%
\right)=\left(%
\begin{array}{ccc}
  m_u & 0 & 0 \\
  0 & m_c & 0 \\
  0 & 0 & m_t \\
\end{array}%
\right), \crn
V^{d+}_L M_d V^d_R&=&
\left(%
\begin{array}{ccc}
  \sqrt{3}h^d_1 u & 0 & 0 \\
  0 & \sqrt{3}h^d_2 u& 0 \\
  0 & 0 & \sqrt{3}h^d_3 v \\
\end{array}%
\right)
=\left(%
\begin{array}{ccc}
  m_d & 0 & 0 \\
  0 & m_s & 0 \\
  0 & 0 & m_b \\
\end{array}%
\right).\label{quarkmasses}\eea

where \bea V^u_R=V^d_R=
\fr{1}{\sqrt{3}}\left(%
\begin{array}{ccc}
  1 & 1 & 1 \\
  \om & \om^2 & 1 \\
  \om^2 & \om & 1 \\
\end{array}%
\right).\eea
The unitary matrices, which couple the left-handed
up- and down -quarks to those in the mass bases, are $U^u_L=1$ and
$U^d_L=1$, respectively. Therefore we get the
CKM matrix at the tree level: \bea
U_\mathrm{CKM}=U^{d\dagger}_L U^u_L=1.\label{a41}\eea
This is a good approximation for the realistic quark mixing
matrix, which implies that the mixings among the quarks
are dynamically small.
The current mass values for the quarks are given by \cite{PDG2014}
\bea m_u&=&(1.8\div3.0)\ \textrm{MeV},\hs m_d=(4.5\div 5.3)\
\textrm{MeV},\hs m_c=(1.25\div1.30)\ \textrm{GeV},\label{vien3}\\
m_s&=&(90.0\div100.0)\ \textrm{MeV},\,\, m_t=(171.99\div 174.43)\
\textrm{GeV},\,\, m_b=(4.15\div4.21)\
\textrm{GeV}.\nn\eea
It is obvious that if $|u| \sim|v|$, the quark
Yukawa couplings can be evaluated from (\ref{quarkmasses}) and (\ref{vien3}) which is the same as in \cite{vlT7}.
We note that with the alignments of $\chi, \eta$ as in (\ref{vevchieta}), $T_{13}\rightarrow Z_3$, i.e, in quark sector, there remains an residual symmetry $Z_3$. If $Z_3$ is broken in to $\{\mathrm{Identity}\}$ by another $SU(3)_L$ multiplet, there will be a contribution for quark sector. A detailed study on this problem has been studied in \cite{vlA4}, so we will not discuss it further.

 \section{\label{neutrino} Neutrino mass and mixing}

The neutrino masses arise from the couplings of $\bar{\psi}^c_{L} \psi_{L}$ to scalars, where $\bar{\psi}^c_{L} \psi_{L}$ transforms as $3^*\oplus 6$ under
$\mathrm{SU}(3)_L$ and $\underline{\bar{3}}_1\oplus \underline{\bar{3}}_1\oplus \underline{3}_2$ under $T_{13}$. For
the known scalar triplets $(\phi, \chi,\eta)$, the
available interactions are only $(\bar{\psi}^c_{i L} \psi_{i
L})\phi$ but
explicitly suppressed because of the $\mathcal{L}$-symmetry. We
will therefore propose new SU(3)$_L$ antisextets, lying in either $ \underline{3}_1$ or $\underline{\bar{3}}_2$ under $T_{13}$,
interact with  $\bar{\psi}^c_{ L}\psi_{ L}$ to produce masses for the neutrino. The antisextet transforms as $s=(s_1,s_2,s_3)\sim \underline{\bar{3}}_2$ under $T_{13}$, with
\bea s_i &=&
\left(%
\begin{array}{ccc}
  s^0_{11} & s^+_{12} & s^0_{13} \\
  s^+_{12} & s^{++}_{22} & s^+_{23} \\
  s^0_{13} & s^+_{23} & s^0_{33} \\
\end{array}%
\right)_i \sim [6^*,2/3,-4/3], (i=1,2,3)\label{sis}\eea where the
numbered subscripts on the component scalars are the
$\mathrm{SU}(3)_L$ indices, whereas $i=1,2,3$ is that of $T_{13}$.

The
VEV of $s$ is set as $(\langle s_1\rangle,\langle s_2,\langle s_3\rangle)$
under $T_{13}$, in which \bea
\langle s_i\rangle=\left(%
\begin{array}{ccc}
  \la_{i } & 0 & v_{i} \\
  0 & 0 & 0 \\
  v_{i} & 0 & \Lambda_{i } \\
\end{array}%
\right). \label{s1}\eea

Following the potential minimization
conditions, we have the followings alignments:
\begin{itemize}
\item[(1)] The first alignment: $\langle s_1\rangle\neq\langle s_2\rangle\neq\langle s_3\rangle$ then $T_{13}$ is completely broken.
\item[(2)] The second alignment: $0\neq\langle s_1\rangle\neq\langle s_2\rangle=\langle s_3\rangle\neq0$ or $0\neq\langle s_1\rangle=\langle s_3\rangle\neq\langle s_2\rangle\neq0$ or $0\neq\langle s_1\rangle=\langle s_2\rangle\neq\langle s_3\rangle\neq0$ then $T_{13}$ is completely broken.
\item[(3)] The third alignment: $\langle s_1\rangle=\langle
s_2\rangle =\langle s_3\rangle \neq 0$ then $T_{13}$ is  broken into $Z_3$ that consists of the elements  \{$e, b, b^2$\}.
\item[(4)] The fourth alignment: $0=\langle s_1\rangle\neq\langle s_2\rangle\neq\langle s_3\rangle \neq 0$ or $0=\langle s_2\rangle\neq\langle s_1\rangle\neq\langle s_3\rangle \neq 0$ or $0\neq \langle s_1\rangle\neq\langle s_2\rangle\neq\langle s_3\rangle=0$ then $T_{13}$ is completely broken.
\item[(5)] The fifth alignment: $0=\langle s_1\rangle\neq\langle s_2\rangle=\langle s_3\rangle \neq 0$ or $0=\langle s_2\rangle\neq\langle s_1\rangle=\langle s_3\rangle \neq 0$ or $0\neq \langle s_1\rangle=\langle s_2\rangle\neq\langle s_3\rangle=0$ then $T_{13}$ is completely broken.
\item[(6)] The sixth alignment: $0=\langle s_1\rangle=\langle s_2\rangle\neq\langle s_3\rangle$ or $0=\langle s_1\rangle=\langle s_3\rangle\neq\langle s_2\rangle$ or $0\neq \langle s_1\rangle\neq \langle s_2\rangle=\langle s_3\rangle=0$ then $T_{13}$ is completely broken.
\end{itemize}
In this work we
impose both the breakings $T_{13}\rightarrow Z_3$ and
$Z_3\rightarrow \{\mathrm{identity}\}\footnote{It means $Z_{3}$ is completely broken.}$ (instead of $T_{13}\rightarrow \{\mathrm{identity}\}$) must be taken place in neutrino sector.
The $T_{13}\rightarrow Z_3$ is achieved with the alignment \bea
\langle s_1\rangle=\langle s_1\rangle=\langle s_1\rangle=\langle s\rangle\neq 0,\label{svev0}\eea where
 \bea \langle s\rangle &=&
\left(%
\begin{array}{ccc}
   \la_{s} & 0 & v_s \\
   0 &  0 & 0 \\
   v_s & 0 &  \La_{s} \\
\end{array}%
\right),\label{svev}\eea
or
\bea
\la_1&=&\la_2=\la_3=\la_s,\,\,\,
v_1=v_2=v_3=v_s,\,\,\,
\La_1=\La_2=\La_3=\La_s.\label{svev1}\eea
The direction of the breaking $Z_{3}\rightarrow
\{\mathrm{identity}\}$  (instead of $T_{13}\rightarrow \{\mathrm{identity}\}$) is achieved in the case $\langle s\rangle=(\langle s\rangle,0,0)$ under $T_{13}$.
To achieve the second direction of the breakings $Z_{3}\rightarrow \{\mathrm{Identity}\}$, we additionally introduce either another scalar $SU(3)_L$ anti-sextet or $SU(3)_L$ triplet which both lies in $\underline{3}_1$ under $T_{13}$.
We can therefore understand the misalignment of the
VEVs in neutrino sector as follows. The $T_{13}$ is broken via two stages, the first
stage is $T_{13}\rightarrow Z_3$ and the second stage is
$T_{13}\rightarrow \{\mathrm{identity}\}$. The first stage is achieved by a $\mathrm{SU}(3)_L$ anti-sextet $s$ with the alignment as in (\ref{svev0}). The second stage can be achieved within each
case below:
 \ben
\item A new $\mathrm{SU}(3)_L$ anti-sextet $\si$ lies in $\underline{3}_1$ under $T_{13}$ ,
\bea \si=\left(%
\begin{array}{ccc}
  \si^0_{11} & \si^+_{12} & \si^0_{13} \\
  \si^+_{12} & \si^{++}_{22}& \si^+_{23} \\
  \si^0_{13} & \si^+_{23} & \si^0_{33} \\
\end{array}%
\right)_i \sim [6^*,2/3,-4/3,\underline{3}_1],\eea
with VEVs is given by $\langle \si\rangle=(\langle \si_1\rangle,0,0)$ under $T_{13}$, where
\bea
\, \langle \si_1\rangle=\left(%
\begin{array}{ccc}
  \la_{\si} & 0 & v_{\si} \\
  0 & 0 & 0 \\
  v_{\si} & 0 & \La_{\si} \\
\end{array}%
\right),\,\, \langle \si_2\rangle=\langle \si_3\rangle=0. \label{sivev}\eea
\item Another
antisextet $s'$ lies in $\underline{\bar{3}}_2$ under $T_{13}$, \bea s'_i =
\left(%
\begin{array}{ccc}
  s'^0_{11} & s'^+_{12} & s'^0_{13} \\
  s'^+_{12} & s'^{++}_{22} & s'^+_{23} \\
  s'^0_{13} & s'^+_{23} & s'^0_{33} \\
\end{array}%
\right)_i \sim [6^*,2/3,-4/3,\underline{\bar{3}}_2],\eea
with VEVs is given by $\langle s'\rangle=(0,\langle s'_2\rangle,0)$ under $T_{13}$, where
\bea \langle s'_2\rangle=\left(%
\begin{array}{ccc}
  \la'_{s} & 0 & v'_{s} \\
  0 & 0 & 0 \\
  v'_{s} & 0 & \Lambda'_{s} \\
\end{array}%
\right). \label{spvev} \eea \een
In calculation, combining both cases we have the Yukawa
interactions:
\bea -\mathcal{L}_\nu&=&\fr 1 2 x (\bar{\psi}^c_{L} \psi_{L})_{\underline{3}_2}s+\fr 1 2 y_1 (\bar{\psi}^c_{L} \psi_{L})_{\underline{\bar{3}}_1}\sigma
 \crn
&+&\fr 1 2 y_2 (\bar{\psi}^c_{L} \psi_{L})_{\underline{\bar{3}}_1}\sigma
+\fr 1 2 z (\bar{\psi}^c_{L} \psi_{L})_{\underline{3}_2}s'+H.c.\crn
&=&
 \fr 1 2 x \left[(\bar{\psi}^c_{1L} \psi_{1L})s_1+
( \bar{\psi}^c_{2L} \psi_{2L}) s_2
 +(\bar{\psi}^c_{3L} \psi_{3L})s_3\right]\crn
&+&\fr 1 2 y_1 \left[(\bar{\psi}^c_{2L} \psi_{3L})\si_1+
( \bar{\psi}^c_{3L} \psi_{1L}) \si_2
 +(\bar{\psi}^c_{1L} \psi_{2L})\si_3\right]\crn
&+&\fr 1 2 y_2 \left[(\bar{\psi}^c_{3L} \psi_{2L})\si_1+
( \bar{\psi}^c_{1L} \psi_{3L}) \si_2
 +(\bar{\psi}^c_{2L} \psi_{1L})\si_3\right]\crn
 &+& \fr 1 2 z \left[(\bar{\psi}^c_{1L} \psi_{1L})s'_1+
( \bar{\psi}^c_{2L} \psi_{2L}) s'_2
 +(\bar{\psi}^c_{3L} \psi_{3L})s'_3\right]+H.c.\label{Lnu}
\eea
Institute (\ref{svev}), (\ref{sivev}) and (\ref{spvev}) into (\ref{Lnu}) we obtain the mass Lagrangian for the
neutrinos:
\bea
-\mathcal{L}^{mass}_\nu&=& \fr 1 2 x\left(\la_{s}\bar{\nu}^c_{1L}\nu_{1L}+v_{s}\bar{N}_{1R}\nu_{1L} +v_{s}\bar{\nu}^c_{1L}N^c_{1R}+ \Lambda_{s}\bar{N}_{1R}N^c_{1R}\right.\crn
&+&\left.\la_{s}\bar{\nu}^c_{2L}\nu_{2L}+v_{s}\bar{N}_{2R}\nu_{2L} +v_{s}\bar{\nu}^c_{2L}N^c_{2R}+ \Lambda_{s}\bar{N}_{2R}N^c_{2R}\right.\crn
&+&\left.\la_{s}\bar{\nu}^c_{3L}\nu_{3L}+v_{s}\bar{N}_{3R}\nu_{3L} +v_{s}\bar{\nu}^c_{3L}N^c_{3R}+ \Lambda_{s}\bar{N}_{3R}N^c_{3R}\right)\crn
&+& \fr 1 2 y_1 \left(\la_{\si}\bar{\nu}^c_{2L}\nu_{3L}+v_{\si}\bar{N}_{2R}\nu_{3L} +v_{\si}\bar{\nu}^c_{2L}N^c_{3R}+ \Lambda_{\si}\bar{N}_{2R}N^c_{3R}\right)\crn
&+& \fr 1 2 y_2 \left(\la_{\si}\bar{\nu}^c_{3L}\nu_{2L}+v_{\si}\bar{N}_{3R}\nu_{2L} +v_{\si}\bar{\nu}^c_{3L}N^c_{2R}+ \Lambda_{\si}\bar{N}_{3R}N^c_{2R}\right)\crn
&+&\fr 1 2 z\left(\la'_{s}\bar{\nu}^c_{2L}\nu_{2L}+v'_{s}\bar{N}_{2R}\nu_{2L} +v'_{s}\bar{\nu}^c_{2L}N^c_{2R}+ \La'_{s}\bar{N}_{2R}N^c_{2R}\right)
+H.c.\label{vv}\eea
$-\mathcal{L}^{mass}_\nu$ in (\ref{vv}) can be rewriten in the matrix form:
\bea -\mathcal{L}^{\mathrm{mass}}_\nu=\fr 1 2
\bar{\chi}^c_L M_\nu \chi_L+ H.c.,\label{nm}\eea where
\bea \chi_L&\equiv&
\left(%
\begin{array}{c}
  \nu_L \\
  N^c_R \\
\end{array}%
\right),\hs M_\nu\equiv\left(%
\begin{array}{cc}
  M_L & M^T_D \\
  M_D & M_R \\
\end{array}%
\right),\crn
\nu_L&=&(\nu_{1L},\nu_{2L},\nu_{3L})^T, \,\,\, N_R=(N_{1R},N_{2R},N_{3R})^T, \eea 
and
\bea M_{L,R,D}=\left(%
\begin{array}{ccc}
a_{L,R,D} & 0 & 0 \\
0 & a_{L,R,D}+c_{L,R,D} & b_{L,R,D} \\
0 & b_{L,R,D} & a_{L,R,D} \\
\end{array}%
\right),\eea with
\bea
  a_{L} & =&\la_sx, \hs\hs\,\,\,\,\,\,  a_{D} =v_s x, \hs \hs\hs\,\,  a_{R} =\La_s x, \crn
   c_{L} & =&\la'_s z, \hs\hs\,\,\,\,\,\,  c_{D} =v'_s z, \hs \hs\hs\,\,  c_{R} =\La'_s z, \crn
  b_{L} & =&\frac{\la_\si}{2} (y_1+y_2),\,\,\, b_{D}= \frac{v_\si}{2} (y_1+y_2),\hs b_{R} =\frac{\La_\si}{2} (y_1+y_2). \label{abLDR}\eea
Three observed neutrinos gain masses via a combination of type I
and type II seesaw mechanisms derived from (\ref{nm}) as \bea
M_{\mathrm{eff}}=M_L-M_D^TM_R^{-1}M_D=\left(%
\begin{array}{ccc}
  A & 0 & 0 \\
  0 & B_1 & C \\
  0 & C & B_2 \\
\end{array}%
\right), \label{Mef}\eea
where \bea A&=&a_L-\fr{a^2_D}{a_R},\crn
B_1&=&a_L-\fr{[b^2_D+(a_D+c_D)^2-c_L(a_R+c_R)]a_R}{a_R^2-b^2_R+a_Rc_R}+\fr{b_R[2b_D(a_D+c_D)-b_Rc_L]-b^2_Dc_R}{a_R^2-b^2_R+a_Rc_R},\crn
B_2&=&a_L-\fr{(a^2_D+b^2_D)a_R}{a_R^2-b^2_R+a_Rc_R}-\fr{(2b_Db_R-a_Dc_R)a_D}{a_R^2-b^2_R+a_Rc_R},\crn
C&=&b_L-\fr{(b^2_D+a^2_D+a_Dc_R)b_R}{a_R^2-b^2_R+a_Rc_R}+\fr{(2a_Da_R+a_Dc_R+a_Rc_D)b_D}{a_R^2-b^2_R+a_Rc_R}.\label{AB12C}\eea
The matrix $M_{eff}$ in (\ref{Mef}) has three exact eigenvalues given by 
\bea \la_1&=&A,\crn
\la_{2,3}&=&\fr 1 2 \left(B_1 + B_2 \mp\sqrt{(B_1-B_2)^2+4 C^2}\right),\label{m123}\eea
and the corresponding eigenstates are 
\bea \varphi_1=\left(%
\begin{array}{ccc}
   1 \\
 0  \\
 0 \\
\end{array}%
\right),\,\,\, \varphi_2=\left(%
\begin{array}{ccc}
   0 \\
  \fr{K}{\sqrt{K^2+1}}  \\
 \fr{1}{\sqrt{K^2+1}} \\
\end{array}%
\right),\,\,\, \varphi_3=\left(%
\begin{array}{ccc}
  0  \\
 \fr{1}{\sqrt{K^2+1}}  \\
 -\fr{K}{\sqrt{K^2+1}} \\
\end{array}%
\right),\label{l123}\eea
where
\bea
K&=&\frac{B_1 -B_2 -\sqrt{(B_1-B_2)^2+4 C^2}}{2C}.\label{K}
\eea
Until now the values of neutrino masses (or the
absolute neutrino masses) as well as the mass ordering of
neutrinos are unknown. 
The neutrino mass spectrum can be the normal hierarchy ($
|m_1|\simeq |m_2| < |m_3|$), the inverted hierarchy ($|m_3|<
|m_1|\simeq |m_2|$)
 or nearly degenerate ($|m_1|\simeq |m_2|\simeq |m_3| $). 
An upper bound on the absolute value of
neutrino  mass was found from the analysis of the cosmological
data \cite{Tegmark} 
\be m_i\leq 0.6\, \mathrm{eV},\label{upb} \ee
while the upper limit on the sum of neutrino masses given in \cite{planck}
\be \sum^{3}_{i=1}m_i< 0.23\, \mathrm{eV} \label{upbsum}\ee
 In the case of 3-neutrino mixing, the two possible signs of $\Delta
m^2_{23}$ corresponding to two types of neutrino mass spectrum can
be provided as follows:
\ben \item[$\circ$] Normal hierarchy (NH): $
|m_1|\simeq |m_2| < |m_3|,\,\, \Delta m^2_{32}=m^2_3-m^2_2>0.$
\item [$\circ$] Inverted hierarchy (IH): $|m_3|< |m_1|\simeq |m_2|,\,\, \Delta m^2_{32} =m^2_3-m^2_2<0$.\een
As will be discussed below, the model under consideration can provide both normal and inverted
 mass hierarchy. 

\subsection{Normal hierarchy ($\Delta m^2_{32}> 0$)}
In this case, three neutrino masses are \be m_1=\left|\la_3\right|, \,\,
m_2=\left|\la_1\right|, \,\, m_3=\left|\la_2\right|,\label{numassN} \ee  with $\la_i \,\,\,
(i=1,2,3)$ is defined in (\ref{l123}), and the corresponding
eigenstates put in the neutrino mixing matrix: 
\bea U_\nu=\left(%
\begin{array}{ccc}
  0 & 1 & 0 \\
 \fr{1}{\sqrt{K^2+1}}& 0 &  \fr{K}{\sqrt{K^2+1}}  \\
 -\fr{K}{\sqrt{K^2+1}} & 0 & \fr{1}{\sqrt{K^2+1}} \\
\end{array}%
\right). P, \label{neu1}\eea
where $P=\mathrm{diag}(1,\,\,\,1,\,\,\, i)$.
Combining (\ref{lep}) and (\ref{neu1}) we get the lepton mixing matrix:
\bea U^\dagger_L
U_\nu= \fr{1}{\sqrt{3}}\left(%
\begin{array}{ccc}
  \fr{1-K}{\sqrt{K^2+1}} & 1 &  \fr{1+K}{\sqrt{K^2+1}} \\
 \fr{\om(1-K\om)}{\sqrt{K^2+1}} & 1 &  \fr{\om(\om+K)}{\sqrt{K^2+1}} \\
   \fr{\om(\om-K)}{\sqrt{K^2+1}} & 1 &  \fr{\om(K\om+1)}{\sqrt{K^2+1}} \\
\end{array}%
\right).P.\label{Ulep}\eea
We note that in the case $K=-1$, $ U^\dagger_L U_\nu \equiv U_{\mathrm{HPS}}$ which is given in (\ref{Uhps}). 

The value of the Jarlskog invariant $J_{CP}$, which gives a convention-independent measure of CP violation, is defined from (\ref{Ulep}) as 
\bea
J_{CP} = \mathrm{Im}[U_{21} U_{31}^* U_{22}^* U_{32}]=\frac{\sqrt{3}(K^2 - 1)}{18(K^2+1)}. \label{J1}\eea
Combining (\ref{J1}) with the data in \cite{PDG2014} for Normal Hierarchy, 
\bea
J_{CP}= -0.032 \label{Jn}\eea 
we find the corresponding values of $K$, 
\bea
K&=&-0.708 \hs (\mathrm{NH}),\label{Kn}
\eea 
and the lepton mixing matrices are obtained as
\bea 
U_{lep}&=&\left(%
\begin{array}{ccc}
 0.805             & \frac{1}{\sqrt{3}} &0.138\\
-0.402 + 0.119i&\frac{1}{\sqrt{3}}&0.697 - 0.069i \\
-0.402 - 0.119i& \frac{1}{\sqrt{3}} &-0.697 - 0.069i\\
\end{array}%
\right)\times P,\label{Ulep12}\eea
or
\bea
\left|U_{lep}\right|&=&\left(%
\begin{array}{ccc}
 0.805& 0.577 & 0.138\\
0.420 & 0.577 & 0.700\\
0.420& 0.577 &0.700\\
\end{array}%
\right), \label{Ulepab2}
\eea
In the standard parametrization, the lepton mixing
 matrix ($U_{PMNS}$) can be parametrized as
 \bea
 U_{PMNS} = \left(%
\begin{array}{ccc}
    c_{12} c_{13}     & s_{12} c_{13}                    & s_{13} e^{-i\delta}\\
    -s_{12} c_{23}-c_{12} s_{23} s_{13} e^{i \delta} & c_{12} c_{23}-s_{12} s_{23} s_{13} e^{i \delta} &s_{23} c_{13}\\
    s_{12} s_{23}-c_{12} c_{23} s_{13}e^{i \delta}&-c_{12} s_{23}-s_{12} c_{23} s_{13} e^{i \delta}  & c_{23} c_{13} \\
    \end{array}%
\right). \mathcal{P}, \label{Ulepg}\eea
where $\mathcal{P}=\mathrm{diag}(1, e^{i \alpha}, e^{i \beta})$, and
$c_{ij}=\cos \theta_{ij}$, $s_{ij}=\sin \theta_{ij}$ with
$\theta_{12}$, $\theta_{23}$ and $\theta_{13}$ being the
solar, atmospheric and  reactor angles, respectively.
$\delta= [0, 2\pi]$ is the Dirac CP violation phase while $\alpha$ and
$\beta$ are two Majorana CP violation phases.
Using the parametrization in Eq. (\ref{Ulepg}) we get
\be
J_{CP}=\frac{1}{8}\cos\theta_{13}\sin2\theta_{12}\sin2\theta_{23}\sin2
\theta_{13}\sin\delta.\label{Jp1}
\ee
Combining  (\ref{Jp1}) and (\ref{Jn}) with the data given in (\ref{PDG2014}) yields $\sin\delta_{CP}=-0.924 $, i.e, $ \delta_{CP}=-67.5^\circ$ or $ \delta_{CP}=292.5^\circ $.

From Eqs. (\ref{K}) and (\ref{Kn}) we get
\bea
B_1 &=& B_2 + 0.704429 C. \label{BCn}
\eea 
In normal case, i.e, $\Delta m^2_{32}=m^2_{3}-m^2_{2}> 0$, for the
remaining constraints, taking the central values from the data
in \cite{PDG2014} as shown in (\ref{PDG2014}): $\Delta m^2_{21}=7.53\times 10^{-5}\
\mathrm{eV}^2$ and $\Delta m^2_{32}=2.44\times 10^{-3}\
\mathrm{eV}^2$, with $m_{1,2,3}$ given in Eqs. (\ref{m123}), we get a solution \footnote{In fact, this system of equations has four solutions, however, these equations differ only by the sign of $m_{1,2, 3}$ that it is not appear in the neutrino oscillation experiments. So, here we only consider in
detail the solution in (\ref{B2C})} (in [eV]) 
\bea
B_2 &=&-0.5\sqrt{4A^2-0.0003} -1.41243C,\crn
C&=&0.47160\left(\sqrt{A^2+2.44\times 10^{-3}}-\sqrt{A^2-7.53\times 10^{-5}}\right). \label{B2C}
\eea 
With $B_{1,2}$ and $C$ in Eqs. (\ref{BCn}) and (\ref{B2C}), $m_{1,2, 3}$ depends only on one parameter $A\equiv m_2$, so we will consider $m_{1,3}$ as
functions of $A$. However, to have an explicit hierarchy on
neutrino masses, in the following figures, $m_2$ should be
included. By using the upper bound on the absolute value of
neutrino  mass in (\ref{upb}) we can
 restrict the values of $A$:  $A \leq 0.6\,\mathrm{eV}$. However, in this case, $A \in (0.0087, 0.05)\, \mathrm{eV}$ or $A \in (-0.05, -0.0087)\, \mathrm{eV}$ are good regions of $A$
that can reach the realistic neutrino mass hierarchy. 

\begin{figure}[ht]
\bc
\includegraphics[width=14.0cm, height=6.0cm]{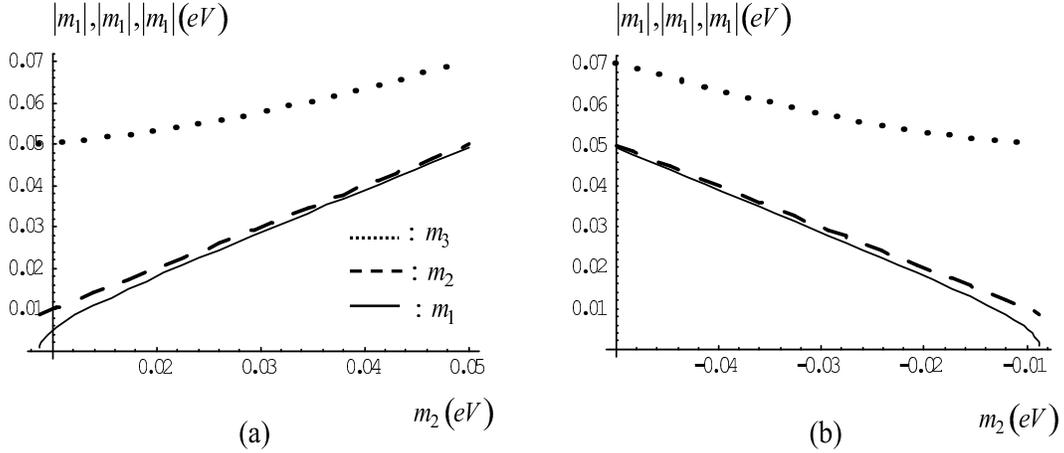}
\vspace*{-0.1cm} \caption[$|m_{1,2, 3}|$ as functions of $A$ in
 the normal hierarchy with a)
 $A\in(0.0087, 0.05) \, \mathrm{eV}$ and b) $A\in(-0.05,-0.0087) \, \mathrm{eV}$.]{$|m_{1,2, 3}|$ as functions of $A$ in
 the normal hierarchy with a)
 $A\in(0.0087, 0.05) \, \mathrm{eV}$ and b) $A\in(-0.05,-0.0087) \, \mathrm{eV}$.}\label{m123N}
\ec
\end{figure}

\begin{figure}[ht]
\bc
\includegraphics[width=7.0cm, height=5.5cm]{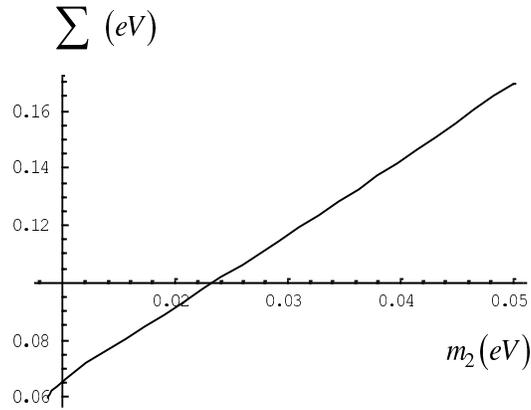}
\vspace*{-0.1cm} \caption[$\sum$ as a function of $A$ with
 $A\in(0.00867, 0.05) \, \mathrm{eV}$ in
 the normal hierarchy.]{$\sum$ as a function of $A$ in
 the normal hierarchy.}\label{m123Ns}
\ec
\end{figure}

In Fig. \ref{m123N}, we have plotted the  absolute value  $|m_{1,2,3}|$
as functions of $A$ with  $A \in (0.0087, 0.05)\, \mathrm{eV}$ and $A \in (-0.05,-0.0087)\, \mathrm{eV}$, respectively. This figure shows
  that there exist allowed regions for values $A$ where either normal
  or quasi-degenerate neutrino masses spectrum is achieved.
  The quasi-degenerate mass hierarchy is obtained when $A$ lies
   in a region\footnote{$A$ increases
  but must be small enough because of the scale of $|m_{1,2,3}|$} [$0.05\,\mathrm{eV}, +\infty$). The normal
  mass hierarchy will be obtained if $A$ takes the values around $(0.0087, 0.05)\,
   \mathrm{eV}$ or $(-0.05, -0.0087)\, \mathrm{eV}$. 
The sum of neutrino masses 
   $\sum=\sum^3_{i=1}|m_i|$ with $A \in (0.0087, 0.05)\,\mathrm{eV}$ is depicted in Fig. \ref{m123Ns} which is consistent with the upper limit given in Eq.(\ref{upbsum}).

\begin{figure}[h]
\begin{center}
\includegraphics[width=7.0cm, height=6.0cm]{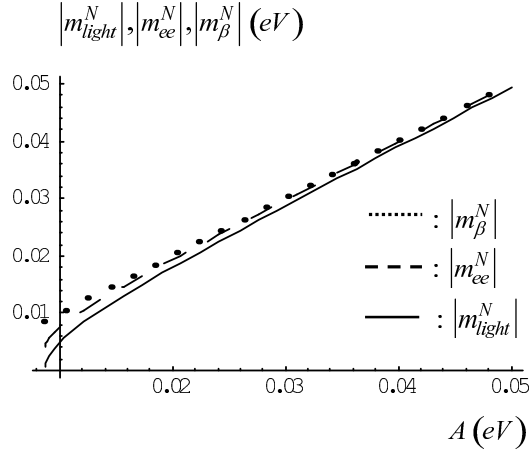}
\caption[$|m^N_{light}|, |m^N_{ee}|$ and $|m^N_{\beta}|$ as functions of $A$ with $A \in (0.0087, 0.05)\,\mathrm{eV}$ in the normal hierarchy]{$|m^N_{light}|, |m^N_{ee}|$ and $|m^N_{\beta}|$ as functions of $A$ with $A \in (0.0087, 0.05)\,\mathrm{eV}$ in the normal hierarchy.}\label{meeNv}
\end{center}
\end{figure}

From the expressions (\ref{K}), (\ref{Ulep}) and (\ref{B2C}), it is easily
 to obtain the effective masses governing neutrinoless double beta decay
 \cite{betdecay1, betdecay2,betdecay3,betdecay4,betdecay5}, 
\bea
m^N_{ee} = \sum^3_{i=1} U_{ei}^2 \left|m_i\right| , \hs
m^N_\beta = \sqrt{\sum^3_{i=1} \left|U_{ei}\right|^2 m_i^2 } \label{meemb} \eea
 which is plotted in Fig. \ref{meeNv} 
with $A \in (0.0087, 0.05)\,\mathrm{eV}$ in the case of $\Delta m^2_{32}> 0$.
We also note
that in the normal spectrum, $|m_1|\approx |m_2|<|m_3|$, so $m_1$
given in (\ref{m123}) is the lightest neutrino mass, which is denoted as $m_{1}\equiv m^N_{light}$.

To get explicit values of the model
parameters, we set $A =10^{-2}\, \mathrm{eV}$, which is safely small.
 Then the other physical neutrino masses are explicitly
 given as
 \bea
 |m_1|\simeq 4.97 \times10^{-3}\, \mathrm{eV},\hs |m_2|=10^{-2}\, \mathrm{eV},\hs |m_3|\simeq 5.04 \times 10^{-2} \, \mathrm{eV}.\eea
 It follows that
\bea
 |m_{ee}^N|&\simeq& 7.51\times10^{-3}\, \mathrm{eV},\hs |m_{\beta}^N|=9.87\times10^{-3}\, \mathrm{eV},\label{meeN}\\
B_1&=&-2.014\times 10^{-2} \,\mathrm{eV},\,\,\, B_2=-3.523\times 10^{-2} \,\mathrm{eV},\,\,\, 
C=2.142\times 10^{-2} \,\mathrm{eV}.\eea
 This solution means a normal
neutrino mass spectrum as mentioned above.
Furthermore, by assuming that\footnote{The values of the parameters $\la_s, \la'_s, \la_\si, v_s, v'_s, v_\si, \La_s, \La'_s, \La_\si$ have not been confirmed by by experiment, however, their hierarchies were given in \cite{dlnvS3, DLpara}. The parametres in Eqs. (\ref{assum}) and (\ref{xyza}) is a set of the model parameters that can fit the experimenta data on neutrino given in (\ref{PDG2014})}
\bea
\la_s=\la'_{s}=\la_{\si}=1\,\mathrm{eV},\,\, v_s=v'_s=v_\si,\,  \La'_s= \La_\si=-\La_s, \La_s=- v^2_s,\label{assum}\eea
 we obtain a solution  
 \bea
 x&=&5\times 10^{-3}, \,\, 
 y_1=y_2\simeq-1.05\times 10^{-2},\,\,
 z\simeq-7.46\times 10^{-3}. \label{xyza}
 \eea
\subsection{Inverted hierarchy ($\Delta m^2_{32}< 0$)}

For inverted hierarchy, three neutrino masses are \be m_1=\left|\la_2\right|, \,\,
m_2=\left|\la_1\right|, \,\, m_3=\left|\la_3\right|,\label{numassI} \ee  with $\la_i \,\,\,
(i=1,2,3)$ is defined in (\ref{l123}), and the corresponding lepton mixing matrix becomes 
\bea U^\dagger_L
U_\nu= \fr{1}{\sqrt{3}}\left(%
\begin{array}{ccc}
  \fr{1+K}{\sqrt{K^2+1}}& 1 &  \fr{1-K}{\sqrt{K^2+1}}  \\
   \fr{\om(\om+K)}{\sqrt{K^2+1}}& 1 & \fr{\om(1-K\om)}{\sqrt{K^2+1}} \\
   \fr{\om(K\om+1)}{\sqrt{K^2+1}} & 1 & \fr{\om(\om-K)}{\sqrt{K^2+1}}  \\
\end{array}%
\right).P.\label{UlepI}\eea
The value of the Jarlskog invariant $J_{CP}$ is defined from (\ref{UlepI}) as 
\bea
J_{CP} =\frac{\sqrt{3}(1-K^2)}{18(1+K^2)}. \label{J2}\eea
In the inverted hierarchy \cite{PDG2014}, $J_{CP}= -0.029$,
we get \bea
K&=&1.365,\label{Ki}\\
B_1 &=& B_2 +0.63213 C ,\label{BCi}
\eea 
and the lepton mixing matrices in (\ref{UlepI}) becomes
\bea 
U^I_{lep}&=&\left(%
\begin{array}{ccc}
 0.807             & \frac{1}{\sqrt{3}} &-0.125\\
-0.403+0.108i &\frac{1}{\sqrt{3}}&0.062 + 0.699i \\
-0.403-0.108i& \frac{1}{\sqrt{3}} &0.062 - 0.699i\\
\end{array}%
\right)\times P.\label{UlepI1}\eea

In the case of the inverted spectrum, combining  (\ref{Jp1})  with the data given in \cite{PDG2014}, $J_{CP}= -0.029$, we get $\sin\delta_{CP}=-0.837136 $, i.e, $ \delta_{CP}=-56.839^\circ$ or $ \delta_{CP}=303.161^\circ $.

Now, by taking the central values from the data
in \cite{PDG2014} as shown in (\ref{PDG2014}): $\Delta m^2_{21}=7.53\times 10^{-5}\
\mathrm{eV}^2$ and $\Delta m^2_{32}=-2.52\times 10^{-3}\
\mathrm{eV}^2$, with $m_{1,2,3}$ is given in  (\ref{numassI}) and $l_{1,2,3}$ in Eq. (\ref{l123}), we get a solution (in [eV]) as follows \footnote{Similarly to in the normal case, there are four solutions in the inverted hierarchy. Here we only consider in
detail the solution in (\ref{BCi})}:
\bea
B_2&=&0.349313\sqrt{A^2-2.52\times 10^{-3}}-0.650687\sqrt{A^2-7.53\times 10^{-5}},\crn
C&=&0.476753\left(\sqrt{A^2-7.53\times 10^{-5}}-\sqrt{A^2-2.52\times 10^{-3}}\right).\label{BCi}
\eea
In this case,
  $A \in (0.05, 0.1)\, \mathrm{eV}$ or $A \in (-0.01, -0.05)\, \mathrm{eV}$ are good regions of $A$
that can reach the realistic neutrino mass hierarchy. Eqs.  (\ref{BCi}) , (\ref{numassI}) and (\ref{l123}) show that $m_{1,2,3}$ only depends on one parameter $A\equiv m_2$, so we will consider $m_{1,3}$ as
functions of $A$. However, to have an explicit hierarchy on
neutrino masses, in the following figures, $m_2$ should be
included.
In Fig. \ref{m123I}, we have plotted the  absolute value  $|m_{1,2,3}|$
as functions of $A$ with  $A \in (0.0502, 0.07)\, \mathrm{eV}$ in the inverted spectrum.

\begin{figure}[ht]
\bc
\includegraphics[width=7.0cm, height=6.5cm]{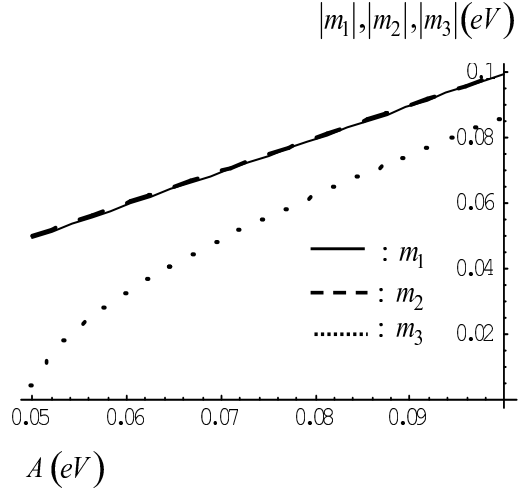}
 \caption[$|m_{1,2,3}|$ as functions of $A$ with
 $A\in(0.05, 0.1) \, \mathrm{eV}$ in
 the inverted hierarchy.]{$|m_{1,2,3}|$ as functions of $A$ with
 $A\in(0.05, 0.1) \, \mathrm{eV}$ in
 the inverted hierarchy.}\label{m123I}
\ec
\end{figure}
\begin{figure}[ht]
\begin{center}
\includegraphics[width=7.0cm, height=6.0cm]{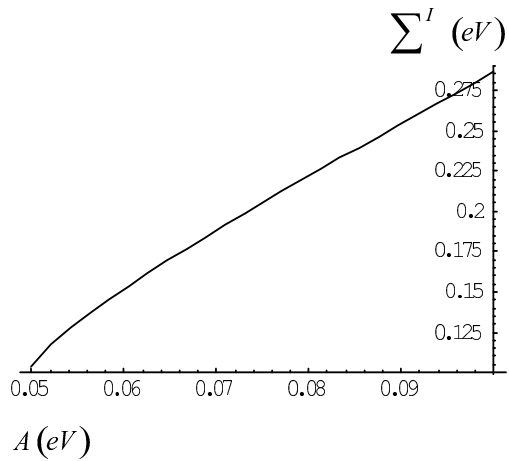}
\caption[ The sum $\sum^I=\sum^3_{i=1}\mid m_i\mid$ as a function of
$A$ with $A \in (0.05, 0.1)\,\mathrm{eV}$ in the case of $\Delta m^2_{32}< 0$]{ The sum $\sum^I=\sum^3_{i=1}\mid m_i\mid$ as a function of
$A$ with $A \in (0.05, 0.1)\,\mathrm{eV}$ in the case of $\Delta m^2_{32}< 0$.}\label{m123Is}
\end{center}
\end{figure}

Fig. \ref{m123I} shows
  that there exist allowed regions for values of $A$ where either inverted
  or quasi-degenerate neutrino masses spectrum is achieved.
  The quasi-degenerate mass hierarchy is obtained if $|A|\in [0.1\,\mathrm{eV} , +\infty$) \footnote{$A$ increases
  but must be small enough because of the scale of $|m_{1,2,3}|$}. The inverted mass hierarchy will be obtained if $A$ takes the values around $(0.05, 0.1)\,
   \mathrm{eV}$ or $(-0.1, -0.05)\, \mathrm{eV}$. The sum
   $\sum^I=\sum^3_{i=1}|m_i|$ with $A \in (0.050, 0.1)\,\mathrm{eV}$ is plotted in the Fig. \ref{m123Is}.

\begin{figure}[ht]
\begin{center}
\includegraphics[width=7.5cm, height=7.0cm]{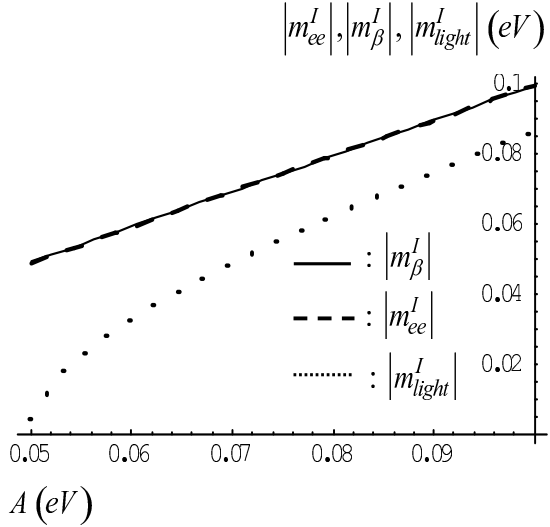}
\caption[$|m^I_{ee}|$,
$|m^I_{\beta}|$ and $|m^I_{light}|$ as functions of $A$ with $A \in (0.05, 0.1)\,\mathrm{eV}$ in the case of $\Delta m^2_{32}<0$]{$|m^I_{ee}|$,
$|m^I_{\beta}|$ and $|m^I_{light}|$ as functions of $A$ with $A \in (0.05, 0.1)\,\mathrm{eV}$ in the case of $\Delta m^2_{32}<0$.}\label{meeI}
\end{center}
\end{figure}

In the inverted spectrum, $|m_3|< |m_1|\approx |m_2|$, so $m_3\equiv m^I_{light}$
given in (\ref{numassI}) is the lightest neutrino mass. In Fig. \ref{meeI}, we have plotted the values of $|m^I_{ee}|$,
$|m^I_{\beta}|$ and $|m^I_{light}|$ as functions of $A$ with $A \in (0.05, 0.1)\,\mathrm{eV}$ in the inverted spectrum.

To get explicit values of the model
parameters, we set $A =.075\, \mathrm{eV}$, which is safely small.
 Then the other physical neutrino masses are explicitly given as
 \bea
 |m_1|=0.0744963\, \mathrm{eV},\hs |m_2|=0.075\, \mathrm{eV},\hs |m_3|=0.0557225 \, \mathrm{eV}.\eea
 It follows that
 $B_1=0.0102354\,\mathrm{eV}$,
 $B_2=-0.0290092\, \mathrm{eV}$ and
$C=0.0620822\, \mathrm{eV}.$
Furthermore, by assuming that
\bea
\la_s=\la'_{s}=\la_{\si}=1\,\mathrm{eV},\,\, v_s=v'_s=v_\si,\, \La_s=a v^2_s, \La'_s=v'^2_s, \La_\si=v^2_\si, \,\, y_1=y_2=y, \label{assumI}\eea
 we obtain a solution as follows
 \bea
 x&=&-0.128008, \hs
 y=-0.135227,\hs
 z=-0.0854825,\hs
 a=0.630556. \label{xyzaI}
 \eea

\section{\label{conclus}Conclusions}

In this paper, we have studied a 3-3-1 model based on non-Abelian discrete symmetry group $T_{13}$ responsible for lepton mixing with non-zero $\theta_{13}$ and CP violation phase. The neutrinos get
small masses and mixing with CP violation phase from $SU(3)_L$ antisextets which are all in triplets under $T_{13}$. If both breakings $T_{13}\rightarrow Z_3$ and $Z_{3}\rightarrow \{\mathrm{Identity}\}$ are taken palce in neutrino sector and $T_{13}$ is broken into $Z_3$ in lepton sector, the realistic neutrino mixing form has been
obtained as a natural consequence of $P_l$ and $T_{13}$ symmetries. The model predicts the lepton mixing with non-zero $\theta_{13}$, and also gives a remarkable prediction of Dirac CP violation $\delta_{CP}=292.5^\circ$ in the normal spectrum, and $\delta_{CP}=303.161^\circ$ in the inverted spectrum which is still missing in the neutrino mixing matrix. There exist some regions of model parameters that can fit the experimental
data in 2014 on neutrino masses and mixing without perturbation. 

\section*{Acknowledgments}
This research is funded by Vietnam National Foundation for Science
and Technology Development (NAFOSTED) under grant number 103.01-2014.51.



\appendix
\section{ $T_{13}$ group theory}
\label{apa}
The $T_{13}$ group is a subgroup of $SU(3)$, and known as the minimal
non-Abelian discrete group having two complex triplets as the irreducible representations.
We denote the generators of $Z_{13}$ and $Z_3$ by $a$ and $b$, respectively.
They satisfy
\begin{equation}
a^{13}=1,\quad ab=ba^9, \quad  b^3=1.
\end{equation}
All of $T_{13}$ elements are written as
\begin{equation}
g=b^{m}a^{n} ,
\end{equation}
with $m=0,1,2$ and $n=0,\cdots,12$.

The generators, $a$ and $b$, are represented e.g. as
\bea a= \left(%
\begin{array}{ccc}
  \rho&0&0\\
  0&\rho^3&0 \\
  0&0&\rho^9\\
\end{array}%
\right), \hs b= \left(%
\begin{array}{ccc}
  0&1&0\\
  0&0&1 \\
  1&0&0\\
\end{array}%
\right).\eea
where $\rho=e^{2i\pi/13}$.
These elements are classified into seven conjugacy classes,
\begin{eqnarray}
\begin{array}{ccc}
 C_1:&\{ e \}, &  h=1,\\
 C_{13}^{(1)}:&\{b~,~ ba~,~ba^{2}~,\quad...\quad,~ ba^{10}~,~ba^{11}~,~ba^{12}\}, &
h=3,\\
  C_{13}^{(2)}:&\{b^2~,~ b^2a~,~b^2a^{2}~,\quad...\quad,~ b^2a^{10}~,~b^2a^{11}~,~b^2a^{12}\}, &  h=3,\\
 C_{3_1}:&\{ a~,~a^{3}~,~a^{9} \},&    h=13,\\
 C_{\bar3_1}:&\{ a^{4}~,~a^{10}~,~a^{12} \},&  h=13,\\
 C_{3_2}:&\{ a^{2}~,~a^{5}~,~a^{6} \},&    h=13,\\
 C_{\bar3_2}:&\{ a^{7}~,~a^{8}~,~a^{11} \},&  h=13.\\
\end{array}
\end{eqnarray}

The $T_{13}$ group has three singlets ${\bf 1}_k$ with $k=0,~1,~2$ and two complex triplets ${\bf 3_1}$ and ${\bf 3_2}$ as irreducible representations.
The characters are shown in Table \ref{T13}, where $\xi_1\equiv \rho+\rho^3+\rho^9$,  $\xi_2\equiv \rho^2+\rho^5+\rho^6$, and $\omega \equiv e^{2i\pi/3}$.

\begin{table}[t]
\begin{center}
\begin{tabular}{|c||c|c|c|c|c|c|c|c|c|}
\hline
        & $n$ &
$h$&$\chi_{\bf 1_{0}}$&$\chi_{\bf 1_{1}}$&$\chi_{\bf1_{2}}$&$\chi_{\bf3_1}$&$\chi_{\bf\bar3_1}$&$\chi_{\bf3_2}$&$\chi_{\bf\bar3_2}$ \\ \hline\hline
$C^{(0)}_1$  & $1$ &$1$&   $1$  &    $1$    &    $1$     &$3$ & $3$  &$3$ & $3$   \\
\hline
$C^{(1)}_{13}$  & $13$ &$3$&   $1$  &    $\omega$    &    $\omega^2$
&$0$&$0$ &$0$&$0$  \\ \hline
$C^{(2)}_{13}$ & $13$  &$3$&   $1$  & $\omega^2$  & $\omega$ &$0$ & $0$  &$0$ & $0$   \\
\hline
$C_{3_1}$ &  $3$  &$13$&   $1$  & $1$  & $1$ &$\xi_1$ &$\bar\xi_1$ &$\xi_2$ &$\bar\xi_2$  \\ \hline
$C_{\bar3_1}$ & $3$ &$13$&   $1$  & $1$&  $1$  &$\bar\xi_1$ &$\xi_1$  &$\bar\xi_2$ &$\xi_2$  \\
\hline
$C_{3_2}$ &  $3$  &$13$&   $1$  & $1$  & $1$ &$\xi_2$ &$\bar\xi_2$ &$\xi_1$ &$\bar\xi_1$  \\ \hline
$C_{\bar3_2}$ & $3$ &$13$&   $1$  & $1$&  $1$  &$\bar\xi_2$ &$\xi_2$  &$\bar\xi_1$ &$\xi_1$  \\
\hline
\end{tabular}
\end{center}
\caption{Characters of $T_{13}$ where $\bar\xi_i \,\, (i=1,2)$ is defined as the complex conjugate of $\xi_i$.}
\label{T13}
\end{table}

Now, let us put $\underline{3}(1,2,3)$ which means some
$\underline{3}$ multiplet such as $x=(x_1,x_2,x_3)\sim
\underline{3}$ or $y=(y_1,y_2,y_3)\sim \underline{3}$ and so on,
and similarly for the other representations. Moreover, the
numbered multiplets such as $(...,ij,...)$ mean $(...,x_i
y_j,...)$ where $x_i$ and $y_j$ are the multiplet components of
different representations $x$ and $y$, respectively. In the
following the components of representations in l.h.s will be
omitted and should be understood, but they always exist in order
in the components of decompositions in r.h.s. All the group multiplication rules of $T_{13}$ as given below.
The tensor products between triplets are obtained as
\bea
\underline{3}_{1}\otimes \underline{3}_{1}&=&\underline{\bar{3}}_{1}(23,31,12)\oplus\underline{\bar{3}}_{1}(32,13,21)\oplus\underline{3}_{2}(11,22,33),\crn
\underline{\bar{3}}_{1}\otimes \underline{\bar{3}}_{1}&=&\underline{3}_{1}(23,31,12)\oplus\underline{3}_{1}(32,13,21)\oplus\underline{\bar{3}}_{2}(11,22,33),\crn
\underline{\bar{3}}_{1}\otimes \underline{3}_{1}&=&\underline{3}_{1}\otimes \underline{\bar{3}}_{1}=1_0(11+22+33)\oplus1_1(11+\om22+\om^233)\crn
&\oplus&1_2(11+\om^222+\om 33)\oplus\underline{3}_{2}(21,32,13)\oplus\underline{\bar{3}}_{2}(12,23,31),\crn
\underline{3}_{2}\otimes \underline{3}_{2}&=&\underline{\bar{3}}_{2}(32,13,21)\oplus\underline{\bar{3}}_{2}(23,31,12)\oplus\underline{\bar{3}}_{1}(22,33,11),\crn
\underline{\bar{3}}_{2}\otimes \underline{\bar{3}}_{2}&=&\underline{3}_{2}(32,13,21)\oplus\underline{3}_{2}(23,31,12)\oplus\underline{3}_{1}(22,33,11),\crn
\underline{\bar{3}}_{2}\otimes \underline{3}_{2}&=&\underline{3}_{2}\otimes \underline{\bar{3}}_{2}=1_0(11+22+33)\oplus1_1(11+\om22+\om^233)\crn
&\oplus&1_2(11+\om^222+\om 33)\oplus\underline{3}_{1}(23,31,12)\oplus\underline{\bar{3}}_{1}(32,13,21),\crn
\underline{3}_{1}\otimes \underline{3}_{2}&=&\underline{3}_{2}(32,13,21)\oplus\underline{\bar{3}}_{2}(31,12,23)\oplus\underline{3}_{1}(33,11,22),\crn
\underline{3}_{1}\otimes \underline{\bar{3}}_{2}&=&\underline{\bar{3}}_{1}(11,22,33)\oplus\underline{\bar{3}}_{2}(23,31,12)\oplus\underline{3}_{1}(21,32,13),\crn
\underline{3}_{2}\otimes \underline{\bar{3}}_{1}&=&\underline{3}_{1}(11,22,33)\oplus\underline{\bar{3}}_{1}(12,23,31)\oplus\underline{3}_{2}(32,13,21),\crn
\underline{\bar{3}}_{1}\otimes \underline{\bar{3}}_{2}&=&\underline{\bar{3}}_{1}(33,11,22)\oplus\underline{\bar{3}}_{2}(32,13,21)\oplus\underline{3}_{2}(31,12,23),\label{12repre}
\eea
The tensor products between singlets are obtained as
\bea
& & \underline{1}_0\otimes \underline{1}_0=\underline{1}_1\otimes \underline{1}_2=\underline{1}_2\otimes \underline{1}_1
=1_0(11),\crn
& & \underline{1}_1\otimes \underline{1}_1=\underline{1}_2(11),
\hs \underline{1}_2\otimes \underline{1}_2=\underline{1}_1(11).\label{11repre}
\eea

The tensor products between triplets and singlets are obtained as
\bea
& &\underline{1}_k\otimes \underline{3}_i=\underline{3}_i(11,12,13),\,\,
\underline{1}_k\otimes \underline{\bar{3}}_i=\underline{\bar{3}}_i(11,12,13) ,\,\, (k=0,1,2; i=1,2).\label{13repre}
\eea

\end{document}